# Imaging Scatterometry

*Morten Hannibal Madsen[1,*], Poul-Erik Hansen[1],*

[1]Danish Fundamental Metrology A/S, Matematiktorvet 307, 2800 Kgs. Lyngby, Denmark

ABSTRACT We present an optical metrology system for characterization of topography of micro/nano-structures on a surface or embedded in a semi-transparent material. Based on the principles of scatterometry, where the intensity of scattered light is used as a 'fingerprint' to reconstruct a surface, this new imaging scatterometer can easily find areas of interest on the cm scale and measure multiple segments simultaneously. The imaging scatterometer measures structural features, such as height, width, and sidewall angle of a grating locally on few µm2 areas with nm resolution. We demonstrate two imaging scatterometers, one built into an optical microscope and one in a split configuration. The two scatterometers are targeted characterization of mm2 and cm2 areas, respectively, and both setups are validated using nano-textured samples.

In recent years, there has been an interest in fast and reliable measurements of micro/nano-textured surfaces. This includes measurements of structural features on surfaces and/or embedded in semi-transparent materials. The interest is mainly driven by the semi-conductor industry for quality control of products, accuracy in production steps and of production equipment. Furthermore, for the decreasing feature sizes and faster production rate in the semiconductor industry, there is a demand for even higher resolution and faster operation of the measurement systems. Other industries and technology areas also show need for such measurement systems, including structural colors[1,2], anti-reflective coatings[3] and hydrophobic surfaces[4]. Many of these new structures are inspired by nature[5] and can now be mass-fabricated using e.g. injection molding[6].

There exist many measurement techniques for dimensional metrology, for example optical microscopy, scanning electron microscopy (SEM), and atomic force microscopy (AFM)[7]. However, these techniques all have one or more disadvantages. Conventional optical microscopes are fast, but are struggling to resolve feature sizes with lateral dimensions less than 1 μm. Advanced optical microscopes, such as interference and confocal microscopes, can improve resolution, but are still limited in their lateral resolution due to the wavelength of the light and high aspect ratio structures with steep slopes are a further challenge[8].

Scanning electron microscopes give high resolution (nm scale) and they are able to provide a large field-of-view (μm to mm scale). However, SEMs require low ambient pressure whereby the dimensions of the sample are limited by the evacuation chamber.

Atomic force microscopy gives high resolution (nm scale), but are hindered by a small field-of-view; typically less than $100\times100$ μm$^2$. Furthermore, AFMs are not able to resolve embedded structures and measurements are laborious and time consuming.

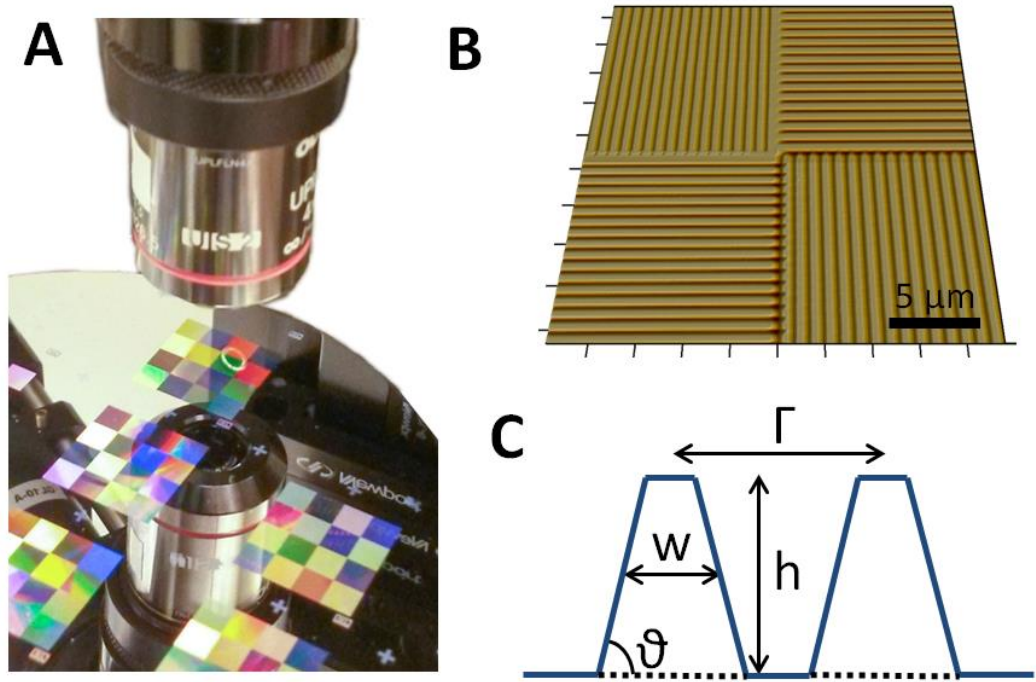

Figure 1. The imaging scatterometry method combines the flexibility of optical microscopy with the resolution of state-of-art microscopes. (A) Visual inspection of a multi-patterned sample with 1D gratings. (B) AFM topography image of the intersection of four areas with different 1D gratings. (C) Sketch of a 1D grating with indication of parameters used in the scatterometry model.

In recent years, scatterometry has become a popular technology for accurate measuring of structural dimensions on a surface or embedded in a semi-transparent layer, as well as of film thicknesses[9–12]. Scatterometry is a characterization technique based on measuring the diffraction intensities of a periodic structured surface, see example of typical structures in Fig. 1B, and compare these to simulated data using an inverse modeling approach. In the simplest approach the comparison is performed by eye[13], but typically diffraction efficiencies are simulated using computer modeling. For the simulation the grating is expressed with typical characteristic parameters such as pitch, height, width, and sidewall angle as indicated in Fig. 1C. Scatterometry provides high resolution (nm scale) similar to SEM and AFM, but at lower cost and more practical handling of samples. However, as scatterometry systems have no information about the scattering from different segments of a sample and are therefore not able to quantify multi-segmented structures.

In this paper we present an imaging scatterometer capable of measuring small areas-of-interest (few μm$^2$) with high resolution (nm scale). Furthermore, the user first has to select the area-of-interest after the images have been acquired and multiple segments can be analyzed within the same image, such as the areas seen in Fig. 1A. The imaging scatterometry method is related to, but simpler to implement than, other techniques such as spectroscopic scatterfield microscopy[14] and imaging ellipsometry[15].

We have designed two experimental setups for imaging scatterometry. The first design is built into a conventional optical microscope (Navitar, 12× zoom) and is ideal for areas up to few mm$^2$. The sample is illuminated using a white LED light source (CLS-LED USB, Qioptiq), which is filtered with a monochromator (CM110, Spectral products) with slit sizes of 0.6 mm, giving a wavelength bandwidth of 6 nm. After the monochromator the light is polarized using a Glan-Laser polarizing crystal. The detector, a monochrome 1.45 megapixel (MPx) charge coupled device (CCD) camera (Pixelink PL-B957), is positioned in the image plane and the sample is brought into focus by adjusting the height of the stage. A 4× infinite focus objective (RMS4X-PF, Olympus) with a numerical aperture of 0.13 is used for the acquisition giving a field of view of around 2 mm$^2$.

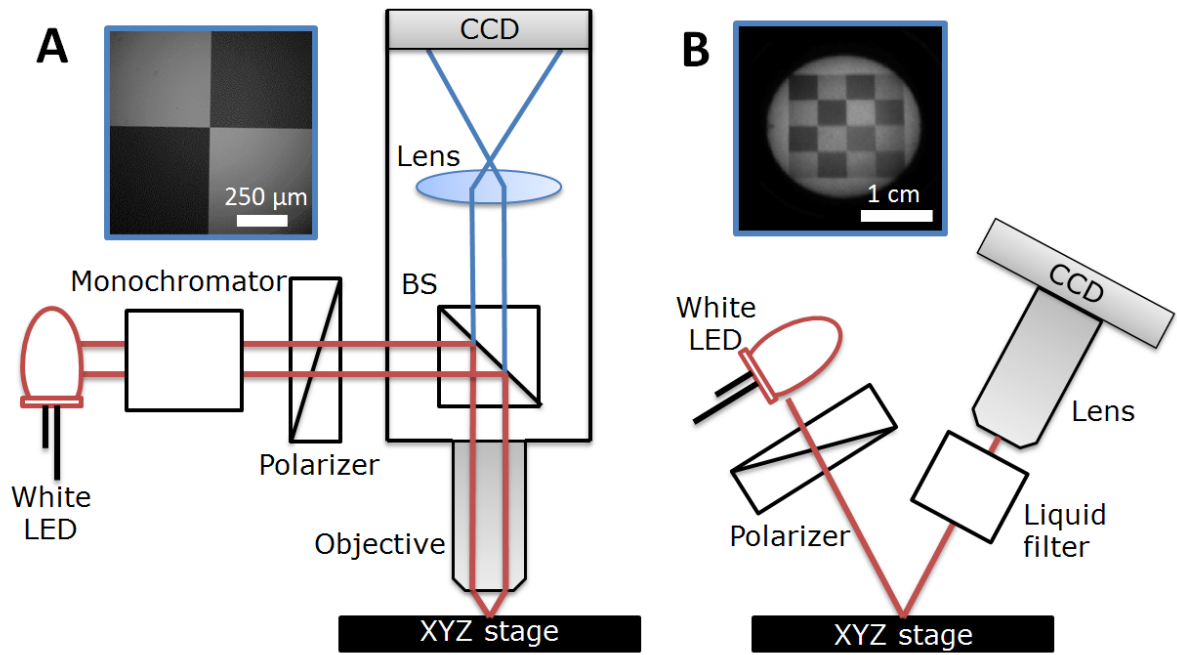

Figure 2. Experimental setups for imaging scatterometry. (A) Sketch of system built into an optical microscope. The light is filtered on the input side using a monochromator. The insert shows the intersection of 4 fields of a multi-structured sample obtained at λ = 540 nm. (B) Sketch of a split configuration system with filtering on the output side. The insert shows 16 fields of a multi-structured sample obtained at λ = 440 nm.

The second design of the imaging scatterometer is a split configuration, where the light source and detector are positioned individually, can be used for areas up to cm$^2$, and even larger areas. With this configuration illumination angles ranging from few degrees to nearly grazing incidence is possible. However, due to the change in focus for depth variations, we limited the angle of incidence to 15 deg. For this setup, the same LED light source is used, but the filtering of the light is performed on the output side using a liquid crystal tunable bandpass filter (Varispec VIS-07-20, PerkinElmer) with an average bandwidth of 10 nm. The light is also polarized using a Glan-Laser polarizer crystal. An 8 mm fixed focal length lens (Navitar, MVL8M23) is mounted in between the tunable filter and the CCD camera. By closing the internal iris of the lens, only nearly parallel light is selected, and thus higher order diffraction peaks is effectively filtered.

To measure the diffraction efficiency, $\eta$, at a given wavelength one should measure the ratio of diffracted light with respect to the incoming light. To also correct for the background signal, e.g. dark noise in the detector and reflectance from optical components, a total of three measurements with the same acquisition settings are necessary for each wavelength. The acquisition time is set to take advantage of the full dynamic range of the camera on the reference area at each wavelength and is for our systems set between 50 ms and 2 s. First a set of reference images, $I_{ref}(\lambda)$, on a surface with known reflection coefficients is acquired, e.g. a Si(100) substrate. Secondly, a set of dark images, $I_{dark}(\lambda)$, is obtained by removing the sample. It is preferable to remove the sample if possible instead of turning the light off, as reflectance from the optical components in the imaging system contributes to the dark signal. And finally a set of images of the structured sample is acquired, $I_{sample}(\lambda)$. The same segments are selected from the three images $I_{ref}(\lambda)$, $I_{dark}(\lambda)$, and $I_{sample}(\lambda)$ leading to segment intensities $I'_{ref}(\lambda)$, $I'_{dark}(\lambda)$, and $I'_{sample}(\lambda)$ from which the diffraction efficiency is found using:

$$\eta(\lambda) = \frac{I'_{sample}(\lambda) - I'_{dark}(\lambda)}{I'_{ref}(\lambda) - I'_{dark}(\lambda)} R(\lambda) \quad , \qquad (0.1)$$

where $R(\lambda)$ is the wavelength dependent reflection coefficients of the reference sample.

We use an inverse modeling approach, where a set of model parameters, *α*, that describes the measured sample is used for simulation of scattering intensities and then compared to the measured data. For this study we have used a 1D grating etched in Si(100) and found that for a given pitch, it can be described by the parameters height, filling factor and sidewall angle, *α* = *α(h,FF,ϑ)*. Fourier modal method algorithms based on the stable implementation of rigorous coupled-wave analysis (RCWA)[16,17] were used for simulation of diffraction efficiencies. The simulated diffraction efficiencies for each set of model parameters were stored in a database and then compared to the experimental data using the chi-square value given by,

$$\chi^2 = \frac{1}{N}\sum_{i=1}^{N}\left[\frac{\eta_i - f(\mathbf{\Omega}_i,\mathbf{\alpha})}{\delta\eta_i}\right]^2 \quad (0.2)$$

where *f*($\mathbf{\Omega}_i$,**α**) are the modelled scattering intensities with the geometry shape parameter **α** and $\mathbf{\Omega}_i$ representing the measuring conditions (wavelength, incident angle and polarization) for the i'th element. *N* is the number of measured diffraction efficiency data points and δ*η* are the experimental uncertainties[17]. The database element with the lowest chi-square value is the best match to the model and the corresponding set of parameters, **α**$_{min}$, gives the best description of the sample. It can be seen from Eq. (0.2) that experimental data points with an associated large uncertainty give a smaller contribution to the sum than data points with a relative small uncertainty. Thus, the best model is found with most weight on the data points with the smallest experimental uncertainties. The confidence limits for the fitting parameters are found using constant chi-square boundaries[18].

As a test sample we used a silicon wafer partly patterned with 1D gratings (16 fields in total, each having a size of 4 mm × 4 mm with pitches ranging from 700 nm to 1400 nm) and large areas without structures as can be seen in Fig. 1(A). The test sample has previously been investigated using state-of-art microscopes. As an example, for the 800 nm pitch area the height has been measured to (189.1 ± 1.3) nm with a metrology AFM, the filling factor to (0.477 ± 0.007) using SEM, and the sidewall angle to (90.1° ± 1.5°) using tilted AFM imaging. The ± denotes the expanded standard uncertainty (k=2) equivalent to a confidence interval of 95 %. A detailed description of these measurements can be found in Ref. [12]. It should be stressed that three separate measurements using different equipment are necessary with the conventional microscopy techniques.

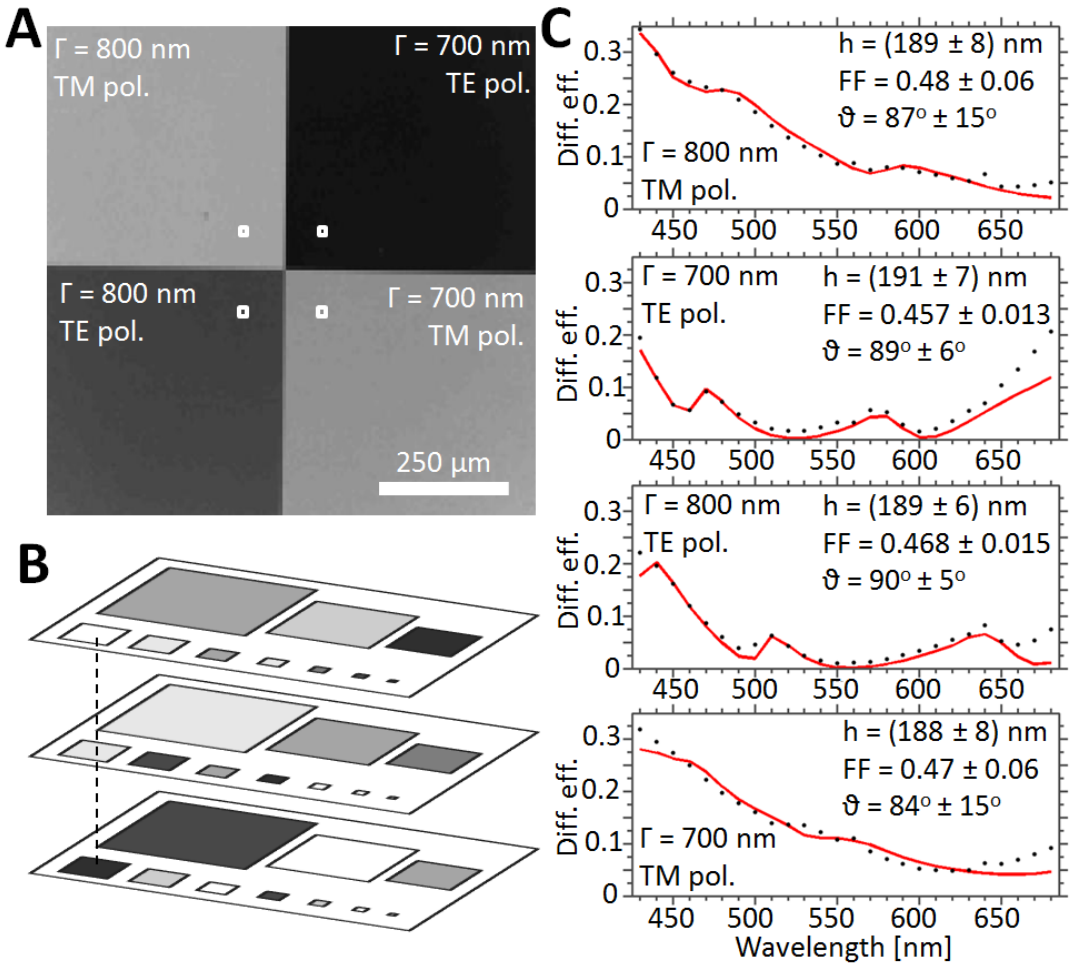

Figure 3. Imaging scatterometry data obtained in the microscope configuration. (A) Pixel based diffraction efficiency image of four areas with different gratings. The analyzed areas are marked with white squares and have the size of (12 × 12) μm². (B) Illustration of the pixel based multi-dimensional analysis of the diffraction efficiency images. (C) Wavelength dependent diffraction efficiencies and best fit for the pixels indicated in (A). The ± denotes the 95 % confidence limits of the fitted parameters.

In the imaging microscope configuration, see Fig. 2A, a set of images that covered the fields with a period of both 700 nm and 800 nm was acquired. A total of 26 images were obtained in the wavelength range from 430 nm to 680 nm to create a multi-spectral image as illustrated in

Fig. 3B. For noise reduction of the images the pixels have been binned together (8×8) resulting in an effective pixels size of (12 × 12) μm$^2$. The wavelength depending diffraction efficiency was calculated for each of these binned pixels using Eq. (0.1).

For each field one pixel was now selected, marked with white squares in Fig. 3A, and the wavelength dependent diffraction efficiencies found for the stack of images as illustrated in Figure 3B. The best fit between measured and simulated diffraction efficiencies was found using Eq. (0.2) and shown in Figure 3C. The incident light polarization is indicated relatively to the orientation of the grating, such that transverse-electric (TE) polarization is parallel with the grating and transverse-magnetic (TM) polarization is perpendicular to the grating. For the given structures there are more minima and maxima in the scattering efficiencies curves for TE polarization than for TM polarization. This makes fitting using the TE polarization more robust and give a more unique solution with narrower confidence intervals.

With the split configuration of the imaging scatterometer, see Fig. 2B, the imaged area is around 2 cm in diameter and a total of 51 images in the wavelength range from 420 nm to 690 nm were acquired. The reference, dark and sample images obtained at 440 nm are shown in Fig. 4(A-C). The calculated diffraction efficiencies at 440 nm for each pixel are shown in Fig. 4D and the wavelength dependent diffraction efficiencies for the field with a pitch of 800 nm, point marked in the image, are shown in Fig. 4E. The reference image is obtained on a flat silicon area in between two areas with nanostructures. Only the area without structures can be used for the reference image, hence the other areas have to be discarded in post-processing. It is a huge advantage that the user first has to discard areas in the post-processing and not while performing the measurements.

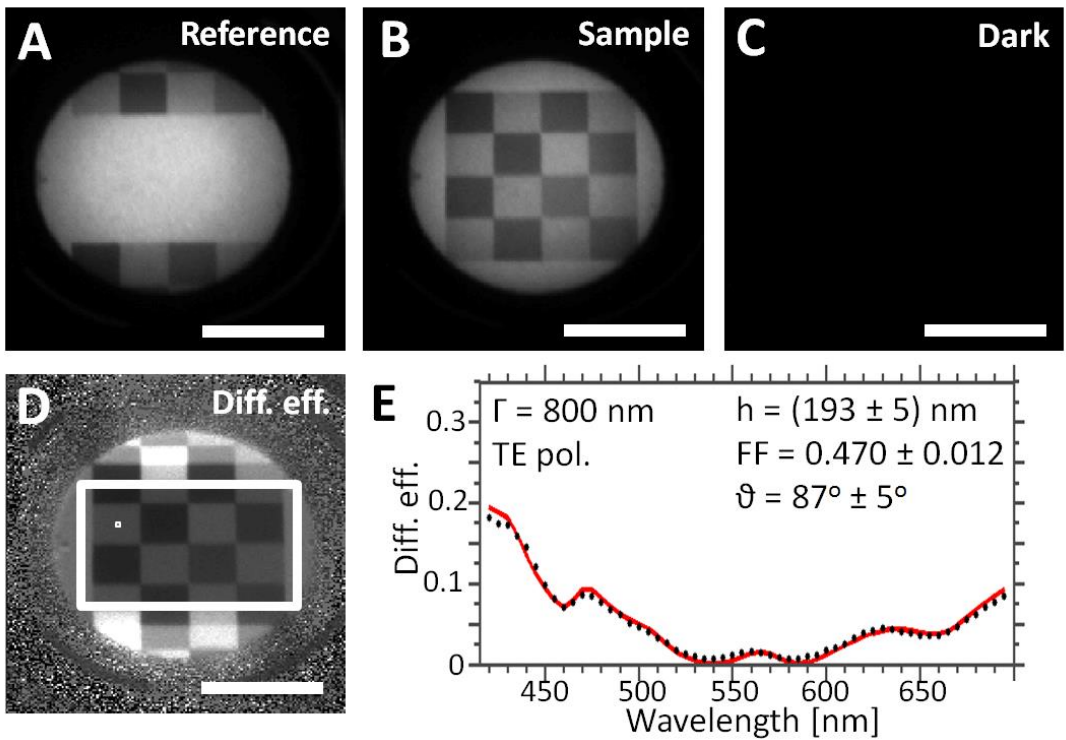

Figure 4. Imaging scatterometry with a split configuration scatterometer. (A) Reference, (B) sample, (C) dark images obtained with the same settings of the camera. (D) Computed diffraction efficiency for each pixel. The pixels have been binned (2 × 2) for noise reduction giving an effective pixel size of (300 × 300) µm$^2$. Only the diffraction efficiencies of the pixels within the white box are valid, as a reference sample with structures outside this area has been used. (E) Wavelength dependent diffraction efficiencies and best fit for the area with a pitch of 800 nm and grating orientation parallel to the polarization of the light. The analyzed pixel is indicated in (D) with a very small white rectangle. Scale bar in (A-D) is 1 cm.

The number of data points is critical for the robustness and uniqueness of the inverse modeling. For both presented methods the acquisition time for a measurement series was around one minute, but the split configuration acquires twice as many images. The acquisition time for known grating structures might be further improved by using a non-linear distribution of wavelengths based on a prior knowledge of minima and maxima positions in the expected line shape spectra.

The measurements are also sensitive to time-dependent fluctuations in the light intensity. Fluctuations faster than 5 ms are not critical as the lowest acquisition time is of the order 25 ms. Fluctuations on a timescale longer than 5 ms and long term drift disturb the measurements. The stability of our light source has been found to be acceptable when operating at full power and long term drift is monitored by obtaining a reference images before and after measuring the sample.

The imaging scatterometry technique is slower compared to a pure spectrometer based setup, but with an image of the surface it is possible to perform a combined defect detection and topography characterization of samples. Conventional scatterometry measures an average of the illuminated area. Defects such as scratches and dust particles disturb the signal and might lead to an offset in the reconstruction. With imaging scatterometry one can avoid areas with defects and thus improve the quality of the measurements. Another important improvement is that areas smaller than the spot size can now be analyzed. This is advantageous, as focusing white light is challenging due to chromatic aberrations.

For all measured samples the three parameters height, width, and sidewall angle the reference measurements are within the confidence interval of the measurements using the imaging scatterometers. However, especially the confidence intervals of the sidewall angle for TM polarized light are large. This is a consequence of scatterometry being a volume sensitive technique, and a change in sidewall angle of a few degrees only have very little influence of the volume of the ridges. The confidence interval is almost the same for the two scatterometer configurations. The noise obtained in the split configuration is slightly higher, but the higher binning of pixels and the increased number of wavelengths that has been measured in the split configuration result in slightly lower confidence limits.

The presented test measurements are on a surface textured sample. However, one of the great advantages of scatterometry is that it can also analyze structures buried in a semi-transparent material, or even under a stack of layers.

In conclusion we have presented a technique for imaging scatterometry and two designs of scatterometers capable of measuring with nm resolution over field of views up to cm$^2$. Imaging scatterometry solves several challenges of non-imaged based scatterometry. These includes an easy method to find the area-of-interest, ensuring that one is measuring inside the same area, that is, fast observation of drift in the system, characterization of multi-segmented samples, and most important, analysis of areas smaller than the spot size. As an example we have shown analysis of areas down to (12 × 12) μm$^2$, but smaller area-of-interest is easily achieved. However, at least a few periods of the grating should be contained within each effective pixel for scatterometry analysis.

**Funding.** The European Union: Theme NMP.2012.1.4-3 (309672).

**Acknowledgment**. We thank NILT for fabrication of the sample with multiple gratings.